%% file: main.tex
\documentclass[runningheads]{main}
\usepackage[T1]{fontenc}

\usepackage{graphicx}
\usepackage[accsupp]{axessibility}
\usepackage{amssymb}
\usepackage{booktabs}
\usepackage{enumitem}
\usepackage{booktabs}
\usepackage{siunitx}
\usepackage[table]{xcolor}
\usepackage{tabularx}
\usepackage{xcolor}
\usepackage{adjustbox}
\usepackage{parskip}
\newif\iffigures
\usepackage[colorlinks,citecolor=blue]{hyperref}
\newcommand{\textbfblue}[1]{\textbf{\textcolor{blue!70!black}{#1}}}
\usepackage{multirow}
\usepackage[capitalize]{cleveref}
\crefname{section}{Sec.}{Secs.}
\Crefname{section}{Section}{Sections}
\Crefname{table}{Table}{Tables}
\crefname{table}{Tab.}{Tabs.}

\definecolor{unet_color}{RGB}{255, 221, 153}   % Faint yellow/orange
\definecolor{ours_color}{RGB}{173, 216, 230}   % Sky blue

\begin{document}

%%%%%%%%% TITLE - PLEASE UPDATE
\title{Harmonized Spatial and Spectral Learning for Generalized Medical Image Segmentation}

\author{Vandan Gorade$^2$, Sparsh Mittal$^1$, Debesh Jha$^2$, Rekha Singhal$^3$, Ulas Bagci$^2$\\ 
$^1$Northwestern University, IL, USA\\
$^2$Indian Institute of Technology, Roorkee, India \\ $^3$TCS research, Mumbai, India
}
% \orcidID{0000-1111-2222-3333} 
\author{
Vandan Gorade\inst{1}, Sparsh Mittal\inst{2}, Debesh Jha\inst{1}, Rekha Singhal\inst{3}, Ulas Bagci\inst{1}
}
\authorrunning{F. Author et al.}
% First names are abbreviated in the running head.
% If there are more than two authors, 'et al.' is used.
%
\institute{ Machine \& Hybrid Intelligence Lab, Northwestern University, Chicago IL, USA \\
\email{{vandan.gorade, debesh.jha, ulas.bagci}@northwestern.edu}
\and ECE Department, Indian Institute of Technology, Roorkee, India \\
\email{sparsh.mittal@ece.iitr.ac.in} \and
TCS Research, Mumbai, India\\
\email{rekha.singhal@tcs.com}
}

\maketitle

\input{sections/titleAbstract}

\input{sections/introduction}

\input{sections/methodology}

\input{sections/ExperimentalPlatform}

\input{sections/results1}

\input{sections/results2}

%%%%%%%%% REFERENCES
{\small
\bibliographystyle{main}
\bibliography{main}
}

\end{document}

%% file: sections/titleAbstract.tex
\begin{abstract}
Deep learning has demonstrated remarkable achievements in medical image segmentation. However, prevailing deep learning models struggle with poor generalization due to  (i) intra-class variations, where the same class appears differently in different samples, and (ii) inter-class independence, resulting in difficulties capturing intricate relationships between distinct objects, leading to higher false negative cases. This paper presents a novel approach that synergies spatial and spectral representations to enhance domain-generalized medical image segmentation. We introduce the innovative Spectral Correlation Coefficient objective to improve the model's capacity to capture middle-order features and contextual long-range dependencies. This objective complements traditional spatial objectives by incorporating valuable spectral information. Extensive experiments reveal that optimizing this objective with existing architectures like UNet and TransUNet significantly enhances generalization, interpretability, and noise robustness, producing more confident predictions. For instance, in cardiac segmentation, we observe a 0.81 pp and 1.63 pp (pp = percentage point) improvement in DSC over UNet and TransUNet, respectively. Our interpretability study demonstrates that, in most tasks, objectives optimized with UNet outperform even TransUNet by introducing global contextual information alongside local details. These findings underscore the versatility and effectiveness of our proposed method across diverse imaging modalities and medical domains. Code is available at \nolinkurl{https://github.com/vangorade/HarmonizedSS_ICPR2024}
% \blfootnote{
% Sparsh and Ulas are the corresponding authors. The computing system used for this research was supported by IIT Roorkee under the grant FIG-100874. The project is supported by the NIH funding: R01-CA246704,  R01-CA240639, U01 DK127384-02S1, and U01-CA268808. Sparsh is supported by the SERB project CRG/2022/003821.}

\end{abstract}

%% file: sections/introduction.tex
%obtaining important anatomical insights, 

\section{Introduction}\label{sec:intro}
\begin{figure}[!htbp]\centering
\def\svgwidth{\columnwidth}
\includegraphics[width=0.85\columnwidth,scale=1.1]{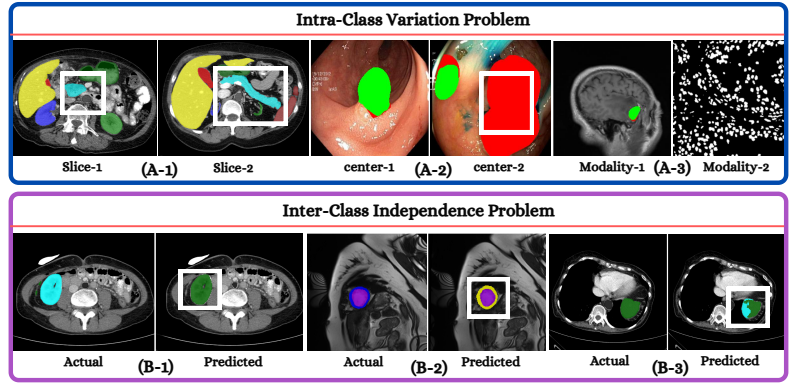}
\caption{(A-1) Appearance disparities within a single class of patient slices, highlighted by white bounding boxes indicating pancreas variation. (A-2) variation in ROI across data acquisition centers. (A-3) ROI variation between modalities. (B-1/2/3) Models face challenges in effectively capturing intricate inter-class relationships, as highlighted by the presence of white bounding boxes. These indicate instances of false negatives, a result of the model's struggle to learn relationships between classes effectively.}
\label{fig: inter-intra}
\end{figure}

Medical image segmentation (MIS) is crucial for supporting clinicians in identifying injuries, monitoring diseases, and planning treatments. Deep learning models have allowed automated delineation of critical structures and organs, enhancing the precision and efficiency of treatment. However, existing deep learning models for MIS\cite{synergynet,huang2020unet} lack generalization \cite{neyshabur2017exploring,kawaguchi2017generalization}, i.e., they fail to accurately segment new and unseen data. 
The challenge to generalization includes the diversity in medical imaging data stemming from variations in imaging devices, protocols, patient demographics, and even the inherent biases \cite{park2022vision,morrison2021exploring,pacl,wang2023theoretical} present in deep learning models. 
%These challenges can be mitigated by tackling the following two issues: 
The diversity manifests as intra-class variations or inter-class independence. (1) Fig.~\ref{fig: inter-intra}(a) depicts  \emph{intra-class variations}. It refers to 
the differences in appearance (size, shape, location, and texture) within a single class, such as organs like the stomach or polyps, across diverse samples from multiple acquisition equipment. 
%The variations may be in , creating complexities in the precise segmentation of organs in various medical images.

(2) Fig.~\ref{fig: inter-intra}(b) shows the \emph{inter-class independence}. It   stems from the model's struggle to effectively model the intricate relationships between distinct objects or classes within the data. For instance, accurately segmenting multiple organs in a CT scan requires a deep understanding of their spatial interactions, influencing their appearances and boundaries. Disregarding such inter-class dependencies may lead to increased false negatives and poor generalization.

\begin{figure}[htbp]\centering
\def\svgwidth{\columnwidth}
\includegraphics[width=0.85\columnwidth,scale=1.5]{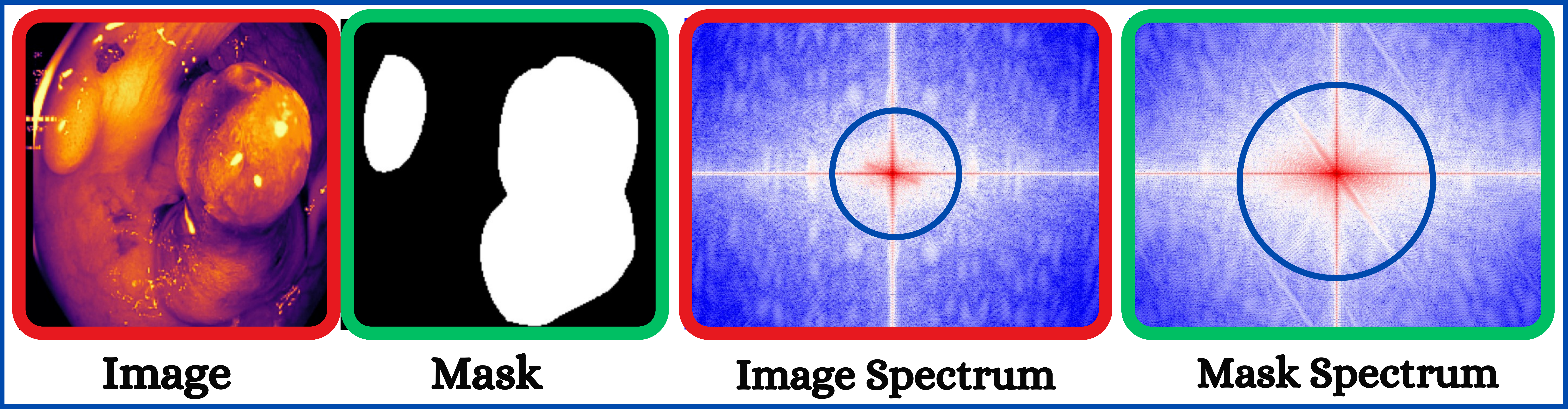}
\caption{A dense low-frequency spectrum (in the middle) indicates that the mask spectrum retains more object information than the image spectrum.}
\label{fig:spectrum_viz}
\end{figure}

We introduce a novel approach that integrates prevalent spatial objectives, such as the Dice Similarity Coefficient, with an innovative objective termed the \emph{Spectral Correlation Coefficient}. Unlike spatial objectives that concentrate on pixel-level comparisons, the Spectral Correlation Coefficient operates in the frequency domain. This integration is intended to augment segmentation models' effectiveness in apprehending middle-order features and contextual long-range dependencies. Both play a vital role in addressing variations within the same class (intra-class variations) and establishing connections between different classes (inter-class dependencies). In contrast to previous methods \cite{yang2022source,yang2020fda,li2023global}, our approach is unique in that it avoids the prevailing practice of applying the Fast Fourier Transform (FFT) to input images. This novelty is important because applying FFT to input images can inadvertently restrict the model's ability to comprehend contextual relationships between objects due to the presence of ROI-irrelevant information in the images, as shown in Fig.~\ref{fig:spectrum_viz}. 

%The unique aspect lies in the 
The Spectral Correlation Coefficient can reveal intricate patterns that remain hidden in the spatial domain. Its computation involves an $\mathcal{O}(N \log N)$ FFT operation, balancing performance with computational overhead.
Our contributions are outlined as follows:

\begin{itemize}
\item We introduce a novel \emph{Spectral Correlation Coefficient} objective, which integrates seamlessly with any architecture. It synergizes spatial and spectral representations, and enables effectively capturing middle-order features and long-range dependencies for domain-generalized MIS.    
     
\item We emphasize that addressing intra-class variations and establishing inter-class dependencies are crucial for achieving domain generalization in medical image segmentation.
        
    \item We conduct experiments on \textbf{\emph{eight}} medical image datasets, comprising diverse imaging modalities and medical domains, e.g., including CT scans, MRIs, skin lesions, histopathology, and polyps. Our method demonstrates significant improvements in segmentation model out-of-distribution (OOD) robustness, enhancing generalization, interpretability, noise resilience, and calibration.   
\end{itemize}

%% file: sections/methodology.tex
\section{Proposed Method}

\subsection{Motivation}

\textbf{Middle-order Features:} Most current segmentation methods rely on spatial objectives to establish correspondence between predicted labels $y$ and the ground truth  $\hat{y}$. However, the raw pixels in the spatial domain exhibit significant noise and often encompass low-order statistics \cite{wei2022masked,bao2021beit}. Transformers and Convolutional Neural Networks (CNNs) possess distinct low-pass and high-pass filtering properties \cite{park2022vision,lsts}, respectively. However, both transformers and CNNs struggle to effectively model certain frequency bands, particularly those related to middle-order features.

Incorporating the medium frequency descriptor, such as the Histogram of Oriented Gradients (HOG), has proven beneficial in enhancing middle-order features \cite{wei2022masked}. This observation has prompted the hypothesis that gaining insights into medium frequencies could potentially aid the model in more effectively learning middle-order features. \textit{Our proposition is that by comprehensively modeling these middle-order features, we can overcome the challenges posed by intra-class variations and inter-class independence.}

\begin{figure}[!t]\centering
\def\svgwidth{\columnwidth}
\includegraphics[width=0.9\columnwidth,scale=1.5]{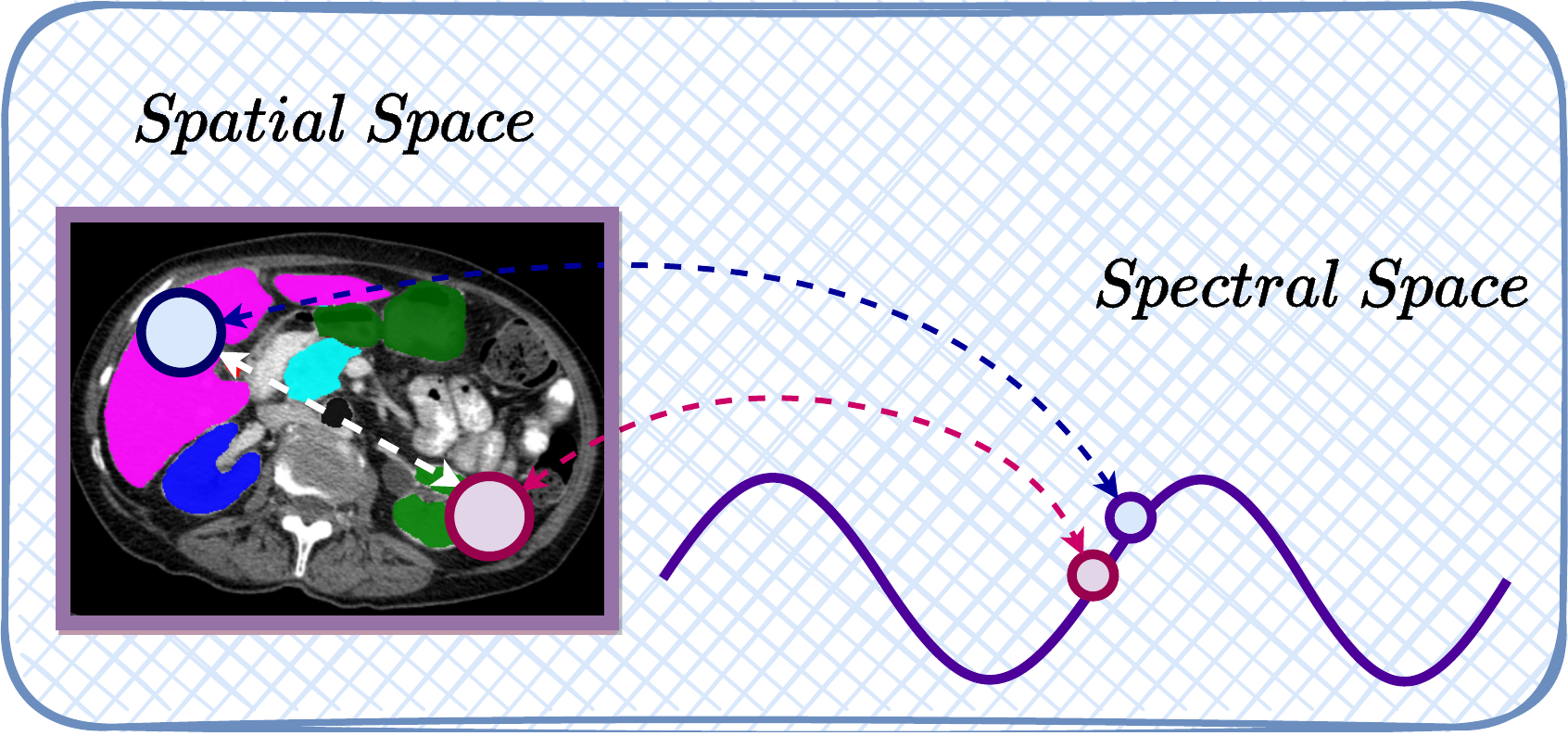}
\caption{Large variations in spatial space correspond to small variations in spectral space and vice versa.}
\label{fig:longrangdep}
\end{figure}

\textbf{Long-range Dependencies:}
Existing CNN architectures face challenges in learning global features \cite{raghu2021vision}, which can lead to difficulties in capturing long-range dependencies. 
In contrast, transformers excel at modeling long-range dependencies \cite{park2022vision}. Nevertheless, we have observed that solely learning long-range dependencies through random patch interactions does not suffice to grasp inter-class dependencies. \textit{We propose that to effectively learn these inter-class dependencies, a model should focus on capturing long-range dependencies between pertinent regions rather than redundant ones.} The frequency space inherently facilitates the modeling of long-range dependencies because minor alterations in frequency space correspond to substantial spatial shifts, as demonstrated in Figure \ref{fig:longrangdep}. With the proposed spectral correlation coefficient, as a model learns correlations between the FFT mask and the predicted mask, it effectively learns the correlations among different frequency components. These components encapsulate only relevant class-related information, allowing us to capture and model inter-class dependencies effectively.

\subsection{Problem Formulation}
% MIS aims to assign accurate labels to anatomical structures. This task is underpinned by a mapping function \( f \), which assigns labels \( y \) to pixels \( x \) in the input image domain \( X \). The overarching goal is to optimize the conditional probability \( P(y | x) \) for inferred segmentation labels \( \hat{y} \).

% \ask{Instead of above 3-4 sentences, can we write:}
Medical image segmentation utilize a mapping function \( f \), which assigns labels \( y \) to pixels \( x \), where the inferred segmentation label is \( \hat{y} \).
 % \ask{To save space, remove this line:} The learning process involves optimizing \( f \)'s parameters by assigning accurate labels using a training dataset. 
The loss function typically combines Binary Cross Entropy (BCE) and the Dice similarity coefficient (Dice), which evaluate the correspondence between predicted labels \( y \) and ground truth segmentation \( \hat{y} \):
\begin{equation}
 \label{eq:spatial}  \mathcal{L}_{\text{spatial}} = BCE(y, \hat{y}) + (1 - Dice(y, \hat{y})) 
\end{equation}

%From the standpoint of domain generalization, the aim 

%, enabling accurate segmentation regardless of variations in imaging settings.
%\end{itemize}

%\ask{Instead of "overcome the constraints imposed", can we say "overcome the limitations of"}

\begin{figure}[!t]\centering
\def\svgwidth{\columnwidth}
\includegraphics[width=0.9\columnwidth,scale=1.5]{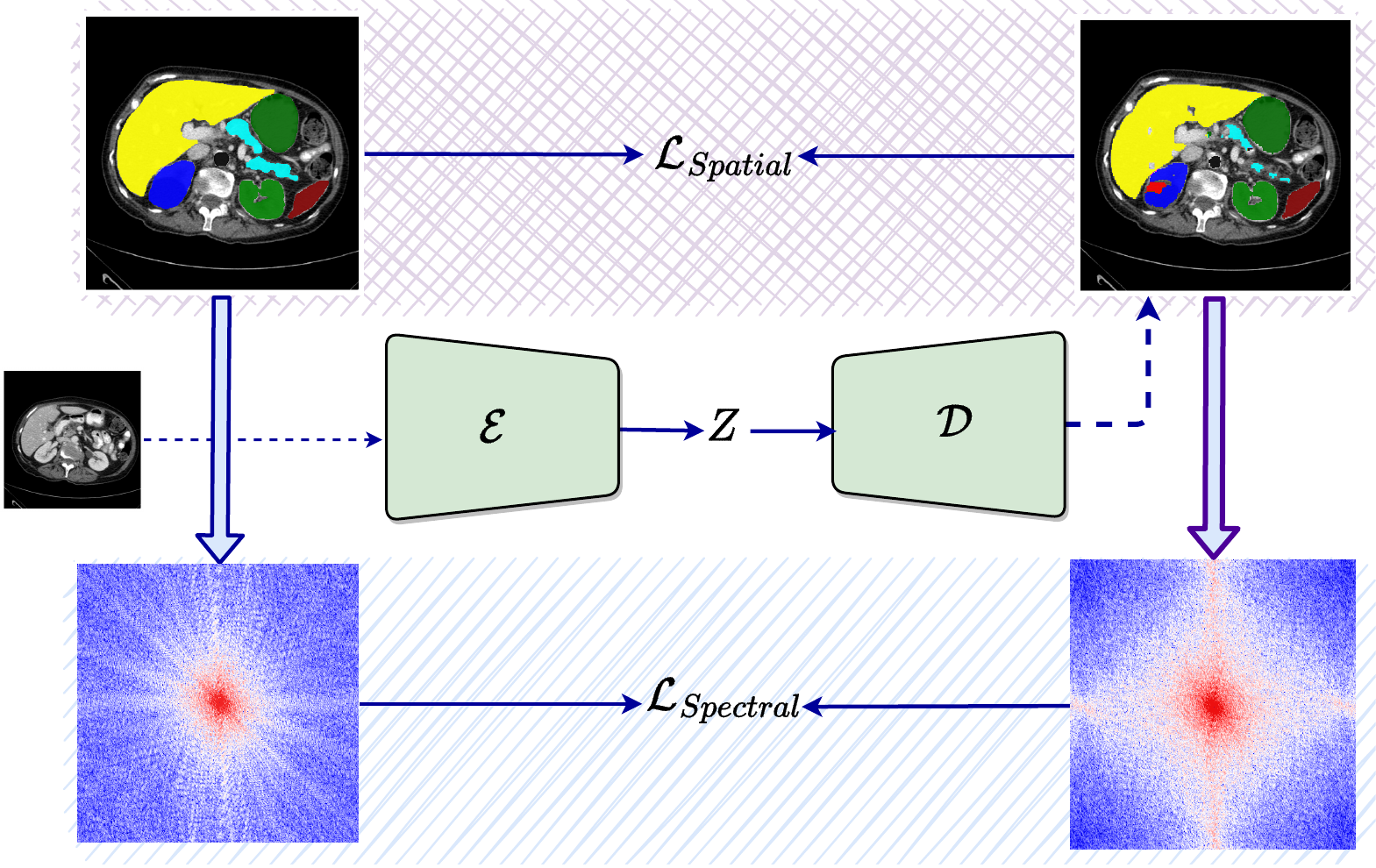}
\caption{Method Workflow: Starting with image $x$ and mask $y$, an encoder-decoder network generates $\hat{y}$. Transforming to spectral space yields $y_{freq}$ and $\hat{y}_{freq}$. Training involves spatial objective $\mathcal{L}_{spatial}$ between $y$ and $\hat{y}$, alongside spectral objective $\mathcal{L}_{spectral}$ between $y_{freq}$ and $\hat{y}_{freq}$.}
\label{fig:workflow}
\end{figure}

%To overcome the constraints imposed by the primary loss function \( \mathcal{L}_{\text{spatial}} \), 
Our goal is to augment \( f \) to transcend specific training domains and generalize effectively across diverse medical image datasets. This entails capturing common features and patterns across different domains. We introduce the Spectral Correlation Coefficient denoted as \( \mathcal{L}_{\text{spectral}} \). This harmonizes the frequency components between predicted and ground-truth masks, effectively mitigating the limitations inherent in \( \mathcal{L}_{\text{spatial}} \). Through the synergistic fusion of \( \mathcal{L}_{\text{spatial}} \) and \( \mathcal{L}_{\text{spectral}} \), the network can more effectively capture intricate inter-class relationships and intra-class variations. This collaborative approach bolsters the model's robustness and efficacy across diverse imaging scenarios. Fig. \ref{fig:workflow} summarizes our approach.

\subsection{Spectral Correlation Coefficient as Regularizer}
Given two spatial binary masks, $y$ and $\hat{y}$, we apply FFT to convert them to the frequency domain. This yields $y_{\text{freq}}$ and $\hat{y}_{\text{freq}}$, which reveal the frequency components inherent to each signal.
Then, we compute the complex inner product between $y_{\text{freq}_i}$ and $\hat{y}_{\text{freq}_i}$ for each index $i$. This complex inner product encapsulates both amplitude and phase interactions in a singular value: $y_{\text{freq}_i} \cdot \overline{\hat{y}_{\text{freq}_i}}$. This helps elucidate the interplay among these frequency components.

By extracting the real component of this complex inner product, denoted as $\text{Re}(y_{\text{freq}_i} \cdot \overline{\hat{y}_{\text{freq}_i}})$, we can discern the interplay between the real and imaginary parts of these frequency components. This reveals the fundamental correlation between them. To measure the strength of these frequency components, we compute the squared magnitude (norm) of each frequency component, yielding $|y_{\text{freq}_i}|^2$ and $|\hat{y}_{\text{freq}_i}|^2$.

These insights culminate in  $\mathcal{L}_{spectral}$, a quantitative metric for correlating $y_{\text{freq}}$ and $\hat{y}_{\text{freq}}$:
\begin{equation}
\label{eq:spectral}
\mathcal{L}_{\text{spectral}} = \frac{2 \sum_{i=1}^N (\text{Re}(y_{\text{freq}_i}) \text{Re}(\hat{y}_{\text{freq}_i}) + \text{Im}(y_{\text{freq}_i}) \text{Im}(\hat{y}_{\text{freq}_i}))}{\sum_{i=1}^N (|y_{\text{freq}_i}|^2 + |\hat{y}_{\text{freq}_i}|^2)}
\end{equation}
Here, $N$ denotes the number of samples in the batch. This equation affords a comprehensive perspective on the similarity between $y_{\text{freq}}$ and $\hat{y}_{\text{freq}}$, effectively encapsulating both their amplitude and phase characteristics. $\mathcal{L}_{spectral}$ stands as a vital metric for quantifying the correlation and shared attributes among frequency components across distinct signals.  
%We synergize \ref{eq:spatial} and \ref{eq:spectral} to obtain the
Our final loss function, $\mathcal{L}_{final} = \mathcal{L}_{spatial} + \lambda \times \mathcal{L}_{spectral}$.  synergizes the complementary representations of individual loss functions.  Here, $\lambda$ is a hyperparameter for smoothly interpolating between spatial and spectral representation. 
Please refer to supplementary material for sensitivity analysis of $\lambda$.

%The terms \( BCE(y, \hat{y}) \) and \( Dice(y, \hat{y}) \) quantify the binary cross-entropy loss and the dice similarity coefficient, respectively, evaluating the correspondence between predicted labels \( y \) and ground truth segmentation \( \hat{y} \).
%Given, two spatial binary masks, $y$ and $\hat{y}$, we perform a transformation into the frequency domain, yielding $y_{\text{freq}}$ and $\hat{y}_{\text{freq}}$. This transformation is accomplished via the Fast Fourier Transform (FFT)\cite{brigham1988fast}, revealing the underlying frequency components inherent to each signal.

%To elucidate the interplay among these frequency components, we delve into the intricacies of their complex inner products. Specifically

%% file: sections/ExperimentalPlatform.tex
\section{Experimental Platform}
We conducted experiments on eight open-source MIS datasets to tackle diverse tasks spanning different anatomical structures.
(1) The Synapse Multi-Organ Segmentation dataset \cite{synapse} comprised 30 clinical CT cases, each equipped with annotated segmentation masks for eight distinct abdominal organs. 
We allocated 18 cases for training and 12 cases for testing \cite{chen2021transunet}. (2) The ACDC dataset \cite{acdc} has 100 cardiac MRI exams, with labels for the left ventricle (LV), right ventricle (RV), and myocardium (MYO). The train:validation:test split is 70:10:20 \cite{chen2021transunet}.  
%We partitioned the dataset into 70 training samples, 10 validation samples, and 20 testing samples \cite{chen2021transunet}. 
(3 \& 4) For polyp segmentation, we used  Kvasir-SEG~\cite{jha2020kvasir} and PolypGen dataset~\cite{ali2021polypgen}. Kvasir-SEG, containing 1000 images, was employed for training, with the official split of 880 training images and the remainder for testing. PolypGen has 1537 images. It assessed model performance under an out-of-distribution (OOD) setting. (5 \& 6) For skin lesion segmentation, we use ISIC-18~\cite{isic18} and ISIC-17~\cite{isic18} datasets. We used the same split as the prior work~\cite{azad2019bi,basak2022mfsnet}. The ISIC-17 test dataset~\cite{mendoncca2013ph}, comprising 650 images, served for OOD testing. (7) For nuclei segmentation, we used the MoNuSeg dataset~\cite{kumar2019multi}, which has 30 images for training and 14 for testing. (8)  Brain Tumour Segmentation (BTSeg) dataset \cite{bts} has 3064 T1-weighted contrast-enhanced images, spanning three types of brain tumors with corresponding binary masks. 
%We maintained a rough 80:20 training-to-test ratio. 
The train: test split is roughly 80:20. For all datasets we set $\lambda = 0.2$.  Please refer to supplementary material for sensitivity analysis of $\lambda$.  

\begin{table*}[!ht]
\centering
\caption{Results on the Kvasir-SEG, ISIC-18, MoNuSeg and BTSeg dataset.}
\label{quant:kviser_isic_monu_bt_seg}
\adjustbox{width=1.0\textwidth}{
\begin{tabular}{p{2.6cm} *{14}{c}}
\toprule
\textbf{Method} & \multicolumn{4}{c}{\textbf{Kvasir-SEG}} & \multicolumn{4}{c}{\textbf{ISIC-18}} & \multicolumn{2}{c}{\textbf{MoNuSeg}} & \multicolumn{4}{c}{\textbf{BTSeg}} \\
\cmidrule(lr){2-5} \cmidrule(lr){6-9} \cmidrule(lr){10-11} \cmidrule(lr){12-15}
 & \textbf{DSC} & \textbf{IOU} & \textbf{SE} & \textbf{SP} & \textbf{DSC} & \textbf{IOU} & \textbf{SE} & \textbf{SP} & \textbf{DSC} & \textbf{IOU}  & \textbf{DSC} & \textbf{IOU} & \textbf{SE} & \textbf{SP} \\
\midrule

UNet
& 88.77 & 76.97 & 81.55 & \textbfblue{98.72}  & 
91.53 & 80.34 & \textbfblue{85.97} & 96.65  & 
72.85 & 58.80
& 84.40 & 68.94 & 74.58 & \textbfblue{99.85} \\
TransUNet
& 87.68 & 75.56 & 81.94 & 98.17 
& 90.92 & 79.42 & 86.30 & 95.44 
& 76.92 & 63.01
& 83.15 & 67.69 & 73.63 & 99.84 \\
\midrule
UNet (Ours) &
\textbfblue{89.40} & \textbfblue{77.03} & \textbfblue{83.19} & {98.53}
& \textbfblue{91.74} & \textbfblue{80.51} & {85.44} & \textbfblue{96.93}
& \textbfblue{73.38} & \textbfblue{58.97}
& \textbfblue{85.48} & \textbfblue{69.91} & \textbfblue{75.33} & \textbfblue{99.85} \\

TransUNet(Ours) &
\textbfblue{89.52} & \textbfblue{77.69} & \textbfblue{83.61} & \textbfblue{98.34}
& \textbfblue{92.00} & \textbfblue{80.81} & \textbfblue{87.21} & \textbfblue{95.79}
& \textbfblue{77.66} & \textbfblue{63.63}
& \textbfblue{87.19} & \textbfblue{71.03} & \textbfblue{74.94} & \textbfblue{99.89} \\

\bottomrule
\end{tabular}
}
\end{table*}
\textbf{Metrics:}  We used the Dice Similarity Coefficient (DSC) and the 95th percentile Hausdorff Distance (HD) metrics on the Synapse and ACDC datasets. For the ISIC-18 and BTSeg datasets, we use  Intersection over Union (IOU), DSC, Specificity (SP), Sensitivity (SE), and Accuracy (ACC). We also use the Expected/Mean Calibration Error (ECE/MCE) to assess the calibration. Lower HD, ECE, and MCE values are better, while higher values for other metrics are better.

%Our framework was implemented in PyTorch and trained on RTX 2080 GPUs. The input image size was set to 224 × 224. 
\textbf{Implementation details:}
We used $224\times224$ images and train on RTX 2080 GPUs using Pytorch. 
During training, we used a batch size of 8 and a learning rate of 0.01. The encoder was initialized with weights pre-trained on ImageNet. We utilized the SGD optimizer with a momentum of 0.9 and weight decay of 0.0001. We employed data augmentations, such as flipping and rotating.
%, to increase the diversity of the training data.

\textbf{Techniques for comparison:}  To ensure a comprehensive and fair evaluation, we chose (1) a CNN-based network, namely UNet with ResNet50 pretrained on ImageNet as the encoder. (2) a  transformer-based network, namely  TransUnet. It has a similar configuration as above, except that it has a transformer bottleneck with eight attention heads.
We trained these models both with and without our proposed $\mathcal{L}_{spectral}$ regularization technique.   We refer to UNet optimized using $\mathcal{L}_{spatial}$ as UNet, and the one optimized using $\mathcal{L}_{final}$  as UNet (ours); same for TransUNet and TransUNet (ours). We maintained uniformity in hyperparameters and architectural configurations across all the methods to isolate the effect of the proposed regularization technique. 
%. This approach guarantees that observed performance differences are primarily attributed to the proposed regularization technique.

%\textbf{Architecture configuration:}

%we carefully chose a single CNN-based and a single transformer-based encoder-decoder architecture.  
%In the CNN-based architecture, we opted for the Unet framework. For this, we integrated a pre-trained ResNet50 encoder, which had been trained on the ImageNet dataset. The decoder component of Unet mirrored the depth of the encoder. For the transformer-based architecture, we used TransUnet. It adhered to a similar configuration as above, except that a transformer bottleneck featuring eight attention heads was incorporated.

%In summary, our study harnessed an assortment of medical datasets 

%% file: sections/results1.tex
\section{Experimental Results}

\subsection{Robustness against Intra-class Variations}
We conducted an extensive analysis to assess the effectiveness of our proposed approach in addressing intra-class variation challenges. Table~\ref{quant:kviser_isic_monu_bt_seg} presents a comprehensive comparison of methods, highlighting the outcomes of our study. Our proposed approach demonstrates clear advantages across diverse datasets that exhibit a wide range of anatomical variations. Notably, our method showcases the ability to accurately delineate both small and large anatomical structures while maintaining fine boundaries.

\begin{figure}[htbp]\centering
\def\svgwidth{\columnwidth}
\includegraphics[width=1.0\columnwidth,scale=1.8]{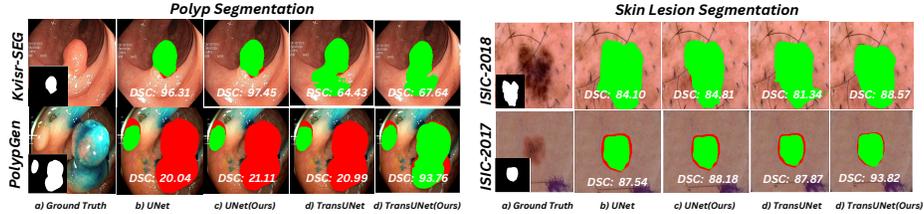}
\caption{Segmentation maps for polyp and skin lesion segmentation: Kvasir-SEG and ISIC-18 are trained under IID settings, while PolypGen and ISIC-17 are treated as OOD datasets. Actual and predicted pathological regions are shown in Red and Green, respectively.}
\label{qual:polyp_skin}
\end{figure}

\textbf{Results on Kvasir-SEG and ISIC-18:} From a quantitative standpoint, on the Kvasir-SEG dataset, UNet(ours) performs comparably or better than the baselines. In fact, the improvements are even more pronounced with TransUNet, with increases of 1.84 pp, 2.13 pp, and 1.67 pp for DSC, IOU, and SE, respectively. The improvement in sensitivity indicates the model's ability to capture positive instances more effectively and reduce false negatives. The ISIC-18 dataset exhibits a similar trend, reaffirming the effectiveness of our approach. A qualitative analysis, as depicted in Fig.~\ref{qual:polyp_skin}, supports our findings. Our proposed method can effectively capture intra-class variations. However, the performance of our method may depend on the nature of the dataset. For instance, datasets such as Kvasir-SEG and ISIC-18 predominantly include segmentation masks with single foreground objects. Such scenarios may limit the effectiveness of our method.

%From a quantitative perspective, our approach consistently yields improvements across both the MoNuSeg and BTSeg datasets, evident through enhanced metrics.

\textbf{Results on MoNuSeg and BTSeg.} Our approach improves the results on both these datasets (Table \ref{quant:kviser_isic_monu_bt_seg}). Specifically, on the MoNuSeg dataset, we observe substantial advancements in performance for both UNet and TransUNet architectures, with increases of 0.53 pp and 0.58 pp in DSC, respectively. Similarly, on the BTSeg dataset, our method significantly benefits both UNet and TransUNet, showcasing DSC improvements of 1.08 pp and 4.04 pp, respectively.
\begin{figure*}
\centering
\def\svgwidth{\columnwidth}
\includegraphics[width=0.9\columnwidth,scale=1.0]{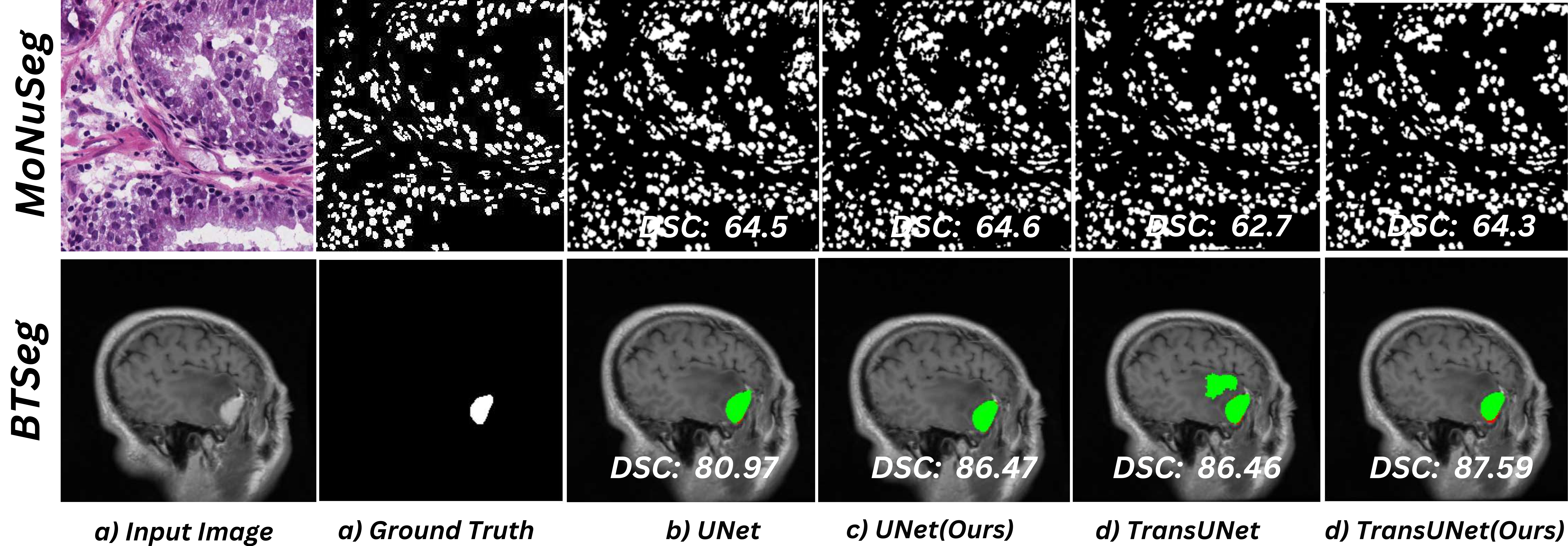}
\caption{Segmentation maps on MoNuSeg and BTSeg. Actual and predicted regions are shown in Red and Green, respectively.}
\label{qual:monu_btseg_pred}
\end{figure*}
Fig~\ref{qual:monu_btseg_pred} offers qualitative insights into our results. Particularly noteworthy is the considerable improvement TransUNet(ours) demonstrated over other baseline methods. This substantial improvement underscores the crucial role of capturing contextual long-range dependencies, achieved through our proposed $\mathcal{L}_{spectral}$ objective. This proves especially advantageous in scenarios like MoNuSeg and BTSeg, where segmentation tasks encompass a wide range of object variations in terms of size, shape, and spatial distribution.

\input{sections/Table1}
\subsection{Robustness against Inter-class Independence}
% #######################################
We comprehensively analyze the effectiveness of our approach in modeling inter-class dependencies. 
The results shown in  Table~\ref{tab:synapse_acdc} distinctly showcase the advantages of synergistically employing both spatial and proposed spectral ($\mathcal{L}_{spectral}$) objectives. Our approach proves to be highly effective in mitigating issues arising from the dependencies between different classes in the segmentation process. Furthermore, it exhibits a clear superiority in accurately delineating both larger, more general objects and intricate fine boundaries between objects.

%From a quantitative perspective, when evaluating the Synapse dataset, 
\begin{figure*}[!ht]\centering
\def\svgwidth{\columnwidth}
\includegraphics[width=1.0\columnwidth,scale=1.8]{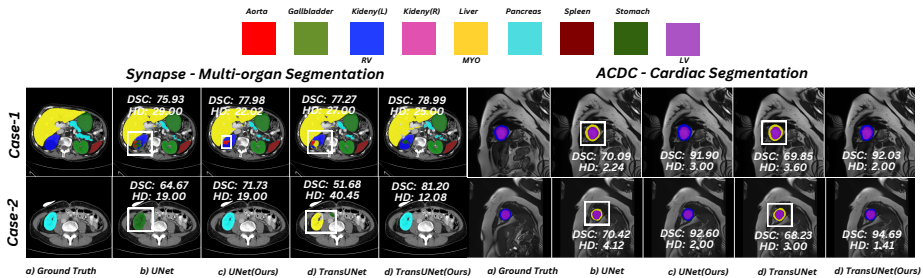}
\caption{Segmentation maps on ACDC and Synapse datasets. The segmentation maps are color-coded to represent different anatomical structures. The overlapping white bounding box represents errors made by the respective model.}
\label{qual:synap_acdc_pred}
\end{figure*}

\textbf{Results on Synapse.} We observed noteworthy improvements in segmentation performance for both UNet (ours) and TransUNet (ours),
compared to UNet and TransUNet. UNet (ours), a CNN architecture, improves DSC by 0.82pp and reduces HD by 5.78pp Interestingly, TransUNet (Ours), a transformer-based architecture, demonstrates a substantial 1.26 pp improvement in DSC, with a surprising increase of 3.18 pp in HD. This suggests that the efficacy of synergizing spatial-spectral representations depends on the specific architecture employed. 

%a CNN's-based architecture, which is optimized using $\mathcal{L}_{final}$ led to a  
%substantial enhancement of 0.82 percentage points (pp) in the mean DSC,  while the mean HD saw a deterioration of 5.48 pp.

The CNNs excel at encoding local information, yet they often struggle to effectively capture global context. In contrast, transformer models are adept at modeling global relationships within data. Our findings reveal that optimizing UNet using $\mathcal{L}_{final}$ led to considerable progress in accurately delineating organs of varying sizes and in capturing intricate fine boundaries. This is due to the spectral regularizer's ability to model contextual long-range dependencies, providing a complementary regularization effect to CNN's strengths. However, TransUNet (ours) tends to overly rely on the global context. While this improved organ delineation, it also led to a limitation in accurately delineating boundaries in multi-organ segmentation scenarios. Fig.~\ref{qual:synap_acdc_pred} further highlights that when equipped with the proposed objective, existing networks can effectively model inter-class dependencies and improved delineation of organs and boundaries.

\textbf{Results on ACDC:} The trends are similar to those on the Synapse dataset. UNet(ours) improves DSC by a  substantial 0.81 pp and reduces HD by 0.44 pp. Interestingly, TransUNet(ours) improves DSC by a substantial 1.63 pp and reduces HD by 0.52 pp. On ACDC dataset, both UNet (ours) and TransUNet (ours) show improved performance in delineating complex multi-scale contour boundaries. The superior performance of TransUNet (ours) suggests that the multi-scale nature of cardiac structure benefits more from the middle to the global context. Fig.~\ref{qual:synap_acdc_pred} highlights that our objective helps existing networks to better model inter-class dependencies and overlapping structures delineation.

%% file: sections/Table1.tex
\begin{table*}[htbp]
\centering
\caption{Results on the Synapse and ACDC dataset. \textbfblue{Blue} indicates the best result.}
\label{tab:synapse_acdc}
\adjustbox{max width=0.95\textwidth}{
\begin{tabular}{p{2.7cm} ccccccccc|ccccc}
\toprule
& \multicolumn{8}{c}{\textbf{Synapse}} & & & \multicolumn{4}{c}{\textbf{ACDC}} \\ 
\cmidrule(lr){1-10} \cmidrule(lr){11-15}
\textbf{Method} & \textbf{Mean} & \multicolumn{8}{c}{\textbf{Class-wise Dice Similarity Coefficient Scores}} & & {\textbf{Mean}} & \multicolumn{3}{c}{\textbf{Class-wise DSC }} \\ 
\cmidrule(lr){1-2} \cmidrule(lr){3-10}  \cmidrule(lr){11-12} \cmidrule(lr){13-15}
 & \textbf{DSC} & \textbf{Aorta} & \textbf{GB} & \textbf{KL} & \textbf{KR} & \textbf{Liver} & \textbf{PC} & \textbf{SP} & \textbf{SM} & &\textbf{DSC} & \textbf{MYO} & \textbf{RL} & \textbf{LV} \\ 
 
\midrule
UNet & 77.54 & 85.52 & \textbfblue{61.86} & 80.57 & 77.24 & 94.37 & 54.72 & \textbfblue{87.95} & 78.12 & & 88.88 & 86.89 & 85.20 & 94.55 \\
TransUNet  &  {77.48}  & \textbfblue{87.23}  & {63.13}  & 81.87  & 77.02  & 94.08  & 55.86  & 85.08  & 75.62 & & 89.69 & {87.42} & 86.80 & 94.88  \\
\midrule
UNet(ours) & \textbfblue{78.36} & \textbfblue{86.42} & {61.16} & \textbfblue{83.55} & \textbfblue{79.64} & \textbfblue{94.44} & \textbfblue{57.69} & {85.67} & \textbfblue{78.32} & & \textbfblue{89.69} & \textbfblue{87.90} & \textbfblue{86.62} & \textbfblue{94.74} \\
TransUNet(ours) & \textbfblue{78.74} & {85.79} & \textbfblue{63.61} & \textbfblue{82.73} & \textbfblue{77.38} & \textbfblue{94.90} & \textbfblue{59.09} & \textbfblue{86.44} & \textbfblue{80.00} & & \textbfblue{91.32} & \textbfblue{90.09} & \textbfblue{88.34} & \textbfblue{95.53} \\
\midrule
 & \textbf{HD} & \multicolumn{8}{c}{\textbf{Class-wise Hausdorff Distance Scores}} & & \textbf{HD} & 
 \multicolumn{3}{c}{\textbf{Class-wise HD}} \\ 
\cmidrule(lr){1-2} \cmidrule(lr){3-10} \cmidrule(lr){11-12} \cmidrule(lr){13-15}
UNet  & 38.26 & 8.06 & 54.21 & \textbfblue{44.52} & 75.69 & 33.67 & 16.92 & \textbfblue{47.81} & 25.17 & & 1.98 & 3.81 & 1.10 & \textbfblue{1.05} \\
TransUNet  & \textbfblue{30.45} & 15.65 & \textbfblue{38.33} & 51.51 & 48.77 & \textbfblue{20.21} & 15.05 & \textbfblue{38.71} & \textbfblue{15.34} & & 1.82 & 3.39 & 1.06 & \textbfblue{1.04}  \\
\midrule
UNet(ours) & \textbfblue{32.48} & \textbfblue{7.17} & \textbfblue{34.37} & {48.99} & \textbfblue{64.63} & \textbfblue{22.09} & \textbfblue{11.82} & {49.36} & \textbfblue{21.42} & & \textbfblue{1.54} & \textbfblue{2.49} & \textbfblue{1.07} & 1.08 \\
TransUNet(ours) & {33.63} & \textbfblue{11.32} & {44.52} & \textbfblue{50.93} & \textbfblue{38.68} & {23.88} & \textbfblue{13.67} & {68.25} & {17.77} & & \textbfblue{1.30} & \textbfblue{1.85} & \textbfblue{1.02} & \textbfblue{1.04} \\
\bottomrule
\end{tabular}
}
\end{table*}

%% file: sections/results2.tex
\subsubsection{Out-of-Distribution Robustness}
Table~\ref{quant:ood_testing} shows the results obtained when models pre-trained on ISIC-18 and Kvasir-SEG are tested on the ISIC-17 and PolypGen datasets.  For the ISIC-17 dataset, both UNet(ours) and TransUNet(ours) demonstrate substantial improvements in both DSC and IOU. TransUNet(ours) is more sensitive compared to others. Moving to the more challenging PolypGen dataset, which comprises polyp data from 6 different centers, we observe a different trend. Specifically, UNet(ours) demonstrates lower generalization capacities compared to UNet. In contrast, TransUNet(ours) exhibits much stronger generalization capabilities. Quantitatively, we observe a 3.72 pp improvement in DSC, a 4.04 pp improvement in IOU, and a 1.49 pp improvement in sensitivity.

\begin{table}[htbp]
\centering
\caption{OOD testing results: \textbf{ISIC-18 $\rightarrow$ ISIC-17} (pre-trained on ISIC-18 and tested on ISIC-17) and \textbf{Kvasir-SEG $\rightarrow$ PolypGen}.}
\label{quant:ood_testing}
\adjustbox{width=0.92\columnwidth}{
\begin{tabular}{p{2.5cm} *{12}{c}}
\toprule
\textbf{Method} & \multicolumn{4}{c}{\textbf{ISIC-18 $\rightarrow$ ISIC-17}} & \multicolumn{4}{c}{\textbf{Kvasir-SEG $\rightarrow$ PolypGen}} \\
\cmidrule(lr){2-5} \cmidrule(lr){6-9}
& \textbf{DSC} & \textbf{IOU} & \textbf{SE} & \textbf{SP} & \textbf{DSC} & \textbf{IOU} & \textbf{SE} & \textbf{SP} \\
\midrule
UNet & 94.01 & 76.65 & 80.50 & \textbfblue{98.25} & \textbfblue{43.98} & \textbfblue{37.15} & \textbfblue{45.77}  & 96.24 \\
TransUNet & 93.61 & 76.41 & \textbfblue{82.84} & 96.85 & 39.99 & 32.92 & 44.14 & 95.05 \\
\midrule
UNet(ours) & \textbfblue{94.38} & \textbfblue{77.20} & \textbfblue{82.05} & 97.84 & 40.35 & 33.81 & 42.06 & \textbfblue{96.27} \\
TransUNet(ours) & \textbfblue{94.86} & \textbfblue{77.84} & 82.62 & \textbfblue{97.78} & \textbfblue{43.71} & \textbfblue{36.96} & \textbfblue{45.63} & \textbfblue{96.63} \\
\bottomrule
\end{tabular}
}
\end{table}

The difference in performance between UNet(ours) and UNet could be attributed to the fact that the proposed objective $\mathcal{L}_{spectral}$, as discussed earlier, is designed to capture relationships and variations between objects present in the mask. However, this dataset may lack such variations or may not have a sufficient amount of them, leading to the observed performance difference. On the other hand, the improved generalization capabilities of TransUNet (ours) can be attributed to the transformer's ability to capture long-range dependencies, in addition to the contribution from middle-order features from $\mathcal{L}_{spectral}$.  Fig~\ref{qual:polyp_skin} provides additional visual evidence of the enhanced capabilities of UNet (ours) and TransUNet (ours) in accurately delineating diverse objects within an out-of-distribution (OOD) setting. These results highlight the versatility of our approach in addressing segmentation challenges across datasets with varying characteristics and complexities. The consistent improvements in performance on both ISIC-17 and PolypGen datasets underscore the generalizability and effectiveness of our proposed method.

\subsection{Calibration Analysis}
We comprehensively evaluate the effectiveness of our proposed approach in generating confident predictions.  Table~\ref{quant:calibration_metrics} shows calibration results for both the In-Distribution (IID) and Out-of-Distribution (OOD) settings. For the IID-setting datasets ISIC-18 and Kvasir-SEG, both UNet(ours) and TransUNet(ours) generate confident predictions compared to their respective baselines. However, on the Kviser dataset, UNet(ours) provides comparable results to UNet, whereas TransUNet(ours) reduces  ECE and MCE by  1.37pp and 2.59pp, respectively. This suggests that under the IID setting, both models consistently provide confident predictions, but their performance varies based on the dataset characteristics. 

\begin{table}[!htbp]
\centering
\caption{Calibration performance under IID and OOD setting.}
\label{quant:calibration_metrics}
\adjustbox{width=0.93\columnwidth}{
\begin{tabular}{p{2.5cm} *{12}{c}}
\toprule
\textbf{Method} & \multicolumn{2}{c}{\textbf{ISIC-18}} & \multicolumn{2}{c}{\textbf{Kvasir-SEG}} & \multicolumn{2}{c}{\textbf{ISIC-18 $\rightarrow$ ISIC-17}} & \multicolumn{2}{c}{\textbf{Kviser $\rightarrow$ PolypGen}} \\
\cmidrule(lr){2-3} \cmidrule(lr){4-5} \cmidrule(lr){6-7} \cmidrule(lr){8-9}
& \textbf{ECE} & \textbf{MCE} & \textbf{ECE} & \textbf{MCE} & \textbf{ECE} & \textbf{MCE} & \textbf{ECE} & \textbf{MCE} \\
\midrule
UNet & 9.13 & 17.51 & 8.61 & \textbfblue{15.85} & \textbfblue{12.80} & \textbfblue{25.00} & {23.67} & {44.53} \\
TransUNet & 9.72 & 18.60 & 9.47 & 17.86 & 14.04 & 27.50 & 27.33 & 51.71 \\
\midrule
UNet(ours) & \textbfblue{8.68} & \textbfblue{16.60} & \textbfblue{8.48} & {16.04} & 13.34 & 26.12 & \textbfblue{23.30} & \textbfblue{43.49} \\
TransUNet(ours) & \textbfblue{9.46} & \textbfblue{18.13} & \textbfblue{8.10} & \textbfblue{15.27} & \textbfblue{13.10} & \textbfblue{25.54}  & \textbfblue{21.19} & \textbfblue{39.58} \\
\bottomrule
\end{tabular}
}
\end{table}

Under the OOD setting, on the ISIC-17 dataset, UNet(ours) generates less confident predictions than UNet, while TransUNet(ours) again shows improved calibration. TransUNet(ours) reduces ECE and MCE by 0.94pp and 1.96pp, respectively. 
For the PolypGen dataset, TransUNet generates highly confident predictions, while UNet exhibits slightly more sensitive behavior and demonstrates slightly improved calibration. In summary, our proposed approach consistently yields confident predictions under both IID and OOD settings.
\subsection{Robustness Against Noise}
%Medical images acquired through 
MRI and CT scan images are often  imperfect due to hardware limitations and patient motion. %Maintaining accurate segmentation amidst such challenges is crucial. 
To test the resilience of our approach, we simulate synthetic Gaussian and Bernoulli noise. Noise levels are set to 0.01, in line with real-world artifacts. As shown in Table~\ref{tab:noise}, UNet (ours) demonstrates improved robustness against noise for the Synapse dataset. However,  TransUNet(ours) loses boundary details (higher HD). For the ACDC dataset, both UNet(ours) and TransUNet(ours) demonstrate improved robustness against noise. The pattern remains same for both the noise types. These results strongly emphasize our proposed objective's efficacy in improving models' generalization under noisy conditions.

\begin{table}[htbp]
\centering
\caption{Performance comparison under noise.}
\label{tab:noise}
\adjustbox{width=0.8\columnwidth}{
\begin{tabular}{p{2.5cm} cccc|cccc}
\toprule
& \multicolumn{4}{c}{\textbf{Gaussian}} &  \multicolumn{4}{c}{\textbf{Bernaulli}} \\
\midrule
\textbf{Method} & \multicolumn{2}{c}{\textbf{Synapse}} & \multicolumn{2}{c}{\textbf{ACDC}} & \multicolumn{2}{c}{\textbf{Synapse}} & \multicolumn{2}{c}{\textbf{ACDC}} \\
\cmidrule(lr){2-3} \cmidrule(lr){4-5} \cmidrule(lr){6-7} \cmidrule(lr){8-9}
 & \textbf{DSC} & \textbf{HD} & \textbf{DSC} & \textbf{HD} & \textbf{DSC} & \textbf{HD} & \textbf{DSC} & \textbf{HD}\\
\midrule
UNet & 70.37 & 37.82 & \textbfblue{73.59} & 3.72 &76.19 & 44.93 & 41.60 & 7.16 \\
TransUNet & 66.59 & \textbfblue{30.15} & 76.62 & 3.04 & 72.02 & \textbfblue{42.68} &49.55 & 5.79 \\
\midrule
UNet(ours) & \textbfblue{71.78} & \textbfblue{29.51} & 73.46 & \textbfblue{2.64} & \textbfblue{77.07} & \textbfblue{32.82} & \textbfblue{46.26} & \textbfblue{6.37} \\
TransUNet(ours) & \textbfblue{70.74} & {36.63} & \textbfblue{79.18} & \textbfblue{2.61} & \textbfblue{76.49} & {49.28} & \textbfblue{53.98} & \textbfblue{5.15} \\
\bottomrule
\end{tabular}
}
\end{table}

\subsection{Interpretability Analysis.}
\textbf{Acquired Qualitative Spectral Maps}
\begin{figure*}[!htbp]\centering
\centering
\def\svgwidth{\columnwidth}
\includegraphics[width=1.0\columnwidth,scale=1.2]{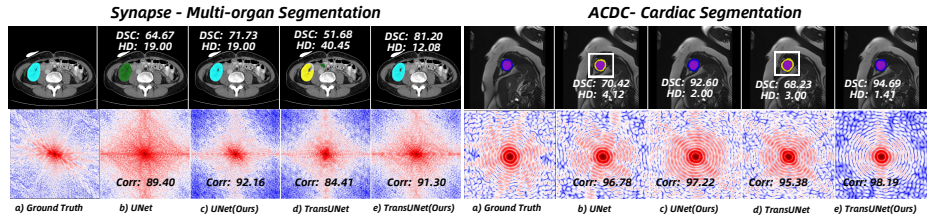}
\caption{Spectral maps for Synapse (row 1 and 2) and ACDC (row 3 and 4) datasets (Corr.= correlation).}
\label{qual:spectral_map}
\end{figure*}
Fig.~\ref{qual:spectral_map} compares each model's spectral maps for Synapse and ACDC datasets. The proposed UNet(ours) and TransUNet(ours) better preserve low to high frequencies compared to the baselines. This improved preservation of spectral information contributes to a higher correlation with the ground-truth spectral map. This makes the predictions more interpretable and aligned with the underlying anatomical structures.

\textbf{Acquired Gradient-weighted Class Activation Maps (CAMs):}
\begin{figure*}[!htbp]\centering
\def\svgwidth{\columnwidth}
\includegraphics[width=0.97\columnwidth,scale=1.3]{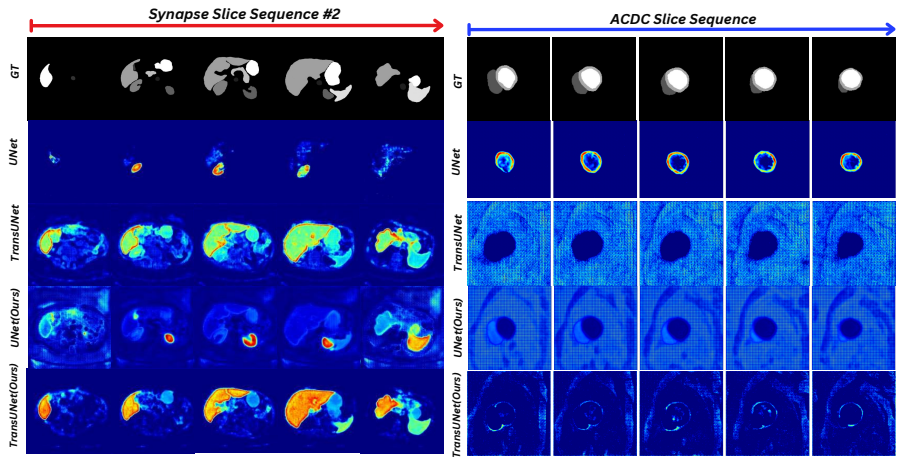}
\caption{Gradient-weighted class activation maps}
\label{qual:CAM_maps}
\end{figure*}
From CAMs provided in Fig~\ref{qual:CAM_maps}, we conclude that: (1) UNet, with its limited receptive field, focuses on local context and overlooks global context, which is crucial for tasks like multi-organ segmentation. Our proposed spectral regularizer enhances UNet's capacity to capture both global contextual relationships while preserving local details across variations. (2) TransUNet tends to emphasize irrelevant regions due to non-contextual long-range dependency modeling. In contrast, our TransUNet(ours) excels at modeling contextual long-range dependencies and middle-order features, thereby attending to both local and global contexts. (3) Our method notably excels in modeling intra-class variations across sequences, surpassing baselines. (4) Unet(ours) offers higher interpretability compared to TransUNet and remains competitive with TransUNet(ours).

\subsection{Sensitivity Analysis}

\begin{table*}[htbp]
\centering
\caption{Sensitivity Analysis}
\label{quant:sensitivity_analysis}
\adjustbox{width=0.5\columnwidth}{
\begin{tabular}{p{1.5cm} c|cc|cc}
\toprule
\multirow{ 2}{*}{\textbf{Backbone}} & \multicolumn{2}{c} {\textbf{ACDC}} & \multicolumn{2}{c}{\textbf{Kviser-SEG}} \\
\cmidrule(lr){2-3} \cmidrule(lr){4-5}
&  \textbf{DSC} & \textbf{HD} & \textbf{DSC} & \textbf{IOU} \\
\midrule
$\lambda$ = 0.1 &  88.96 & 1.88 & 88.35 & 
76.84\\
$\lambda$ = 0.2 &  \textbfblue{89.69} &  \textbfblue{1.54} &  \textbfblue{89.40} & 
77.03\\
$\lambda$ = 0.3 & 89.33 & 1.74 & 88.26 & 
76.84\\
$\lambda$ = 0.5 & 87.54 & 2.10 & 87.34 &   \textbfblue{77.34}\\
$\lambda$ = 0.9 & 79.88 & 3.77 & 84.70 & 73.46 \\
\bottomrule
\end{tabular}}
\end{table*}
In Section 2.3 of the manuscript, we introduce the $\lambda$ hyperparameter, and in Table~\ref{quant:sensitivity_analysis}, we present a comprehensive analysis of its impact on the ACDC and Kviser-SEG datasets. For the ACDC dataset, our analysis reveals that setting $\lambda$ to 0.2 yields the most favorable results in terms of Dice Similarity Coefficient (DSC) and Hausdorff Distance (HD). On the other hand, when considering the Kviser-SEG dataset, we found that a value of $\lambda=0.2$ leads to the highest DSC, whereas $\lambda=0.5$ produces the best Intersection over Union (IOU). Notably, as we increase the value of $\lambda$ beyond these optimal settings, we observe a noticeable degradation in segmentation performance for both datasets. This observation underscores the significance of spatial representation in segmentation tasks, indicating that spatial features offer superior representations. However, it is worth emphasizing that a judiciously crafted weighting scheme that synergizes spatial and spectral information can potentially enhance domain generalization. 

\section{Conclusion, Limitations and Future Work}
In this study, we introduce a novel spectral objective, the spectral correlation coefficient, in synergy with a spatial objective, effectively enhancing domain generalization in medical image segmentation. This approach seamlessly integrates with existing encoder-decoder architectures. When combined with TransUNet, it achieves remarkable performance and outperforms state-of-the-art methods across diverse medical segmentation tasks. Our method exhibits interpretability and resilience to noisy data while generating confident predictions.
Future work will concentrate on minimizing false negatives, especially in noisy environments. One intriguing avenue for future research involves integrating our proposed method into established semi-supervised or knowledge distillation-based\cite{goraderethinking} approaches to increase efficiency in terms of annotation and computation. Additionally, there is potential for extending the scope of our method beyond medical tasks, conducting performance analyses in diverse application domains.

\section{Acknoledgement}
This study was supported by NIH grants R01-CA246704, R01-CA240639, U01 DK127384-02S1, and U01-CA268808. Sparsh is supported by the SERB project CRG/2022/003821. IIT Roorkee provided support for the computing system used for this research under the grant FIG-100874. Sparsh and Ulas are the corresponding authors.
% We introduce a novel spectral objective, the spectral correlation coefficient, synergizing with a spatial objective to enhance domain generalization in MIS. Our proposed objective can be easily integrated with any existing encoder-decoder segmentation architecture. Our thorough empirical analysis showcases proposed method when optimized with TransUNet achieve state-of-art results on variety of medical segmentation tasks. Furthermore, our proposed method is interpretable, robust against noisy data 
% and confident prediction generation. 
% Our thorough empirical analysis showcases the method's enhancements in interpretability, noise robustness, and confident prediction generation. 
% Our future work will focus on further reducing the false negatives, particularly in noisy conditions.

%While the proposed method advances existing models' capabilities, it does not completely reduce false-negative cases, particularly in noisy conditions. This presents a platform for future research endeavors.

%% file: main.bbl
\begin{thebibliography}{10}
\providecommand{\url}[1]{\texttt{#1}}
\providecommand{\urlprefix}{URL }
\providecommand{\doi}[1]{https://doi.org/#1}

\bibitem{synapse}
Multi-atlas abdomen labeling challenge. synapse multi-organ segmentation dataset. Available at: \url{https://www.synapse.org/#!Synapse:syn3193805/wiki/217789} (2015)

\bibitem{acdc}
Acdc (automated cardiac diagnosis challenge). Available at: \url{ https://www.creatis.insa-lyon.fr/Challenge/acdc} (2017)

\bibitem{ali2021polypgen}
Ali, S., Jha, D., Ghatwary, N., Realdon, S., Cannizzaro, R., Salem, O.E., Lamarque, D., Daul, C., Riegler, M.A., Anonsen, K.V., et~al.: A multi-centre polyp detection and segmentation dataset for generalisability assessment. Scientific Data  \textbf{10}(1), ~75 (2023)

\bibitem{azad2019bi}
Azad, R., Asadi-Aghbolaghi, M., Fathy, M., Escalera, S.: Bi-directional convlstm u-net with densley connected convolutions. In: Proceedings of the IEEE/CVF international conference on computer vision workshops. pp.~0--0 (2019)

\bibitem{bao2021beit}
Bao, H., Dong, L., Piao, S., Wei, F.: Beit: Bert pre-training of image transformers. arXiv preprint arXiv:2106.08254  (2021)

\bibitem{basak2022mfsnet}
Basak, H., Kundu, R., Sarkar, R.: Mfsnet: A multi focus segmentation network for skin lesion segmentation. Pattern Recognition  \textbf{128},  108673 (2022)

\bibitem{chen2021transunet}
Chen, J., Lu, Y., Yu, Q., Luo, X., Adeli, E., Wang, Y., Lu, L., Yuille, A.L., Zhou, Y.: Transunet: Transformers make strong encoders for medical image segmentation. arXiv preprint arXiv:2102.04306  (2021)

\bibitem{bts}
Cheng, J.: Brain tumor dataset. \url{ https://doi.org/10.6084/m9.figshare.1512427.v5} (2017)

\bibitem{isic18}
Codella, N.C., Gutman, D., Celebi, M.E., Helba, B., Marchetti, M.A., Dusza, S.W., Kalloo, A., Liopyris, K., Mishra, N., Kittler, H., et~al.: Skin lesion analysis toward melanoma detection: A challenge at the 2017 international symposium on biomedical imaging (isbi), hosted by the international skin imaging collaboration (isic). In: 2018 IEEE 15th international symposium on biomedical imaging (ISBI 2018). pp. 168--172 (2018)

\bibitem{goraderethinking}
Gorade, V., Mittal, S., Jha, D., Bagci, U.: Rethinking intermediate layers design in knowledge distillation for kidney and liver tumor segmentation. ArXiv

\bibitem{synergynet}
Gorade, V., Mittal, S., Jha, D., Bagci, U.: Synergynet: Bridging the gap between discrete and continuous representations for precise medical image segmentation. In: Proceedings of the IEEE/CVF Winter Conference on Applications of Computer Vision. pp. 7768--7777 (2024)

\bibitem{pacl}
Gorade, V., Mittal, S., Singhal, R.: Pacl: Patient-aware contrastive learning through metadata refinement for generalized early disease diagnosis. Computers in Biology and Medicine  \textbf{167},  107569 (2023)

\bibitem{lsts}
Gorade, V., Singh, A., Mishra, D.: Large scale time-series representation learning via simultaneous low-and high-frequency feature bootstrapping. IEEE Transactions on Neural Networks and Learning Systems  (2023)

\bibitem{huang2020unet}
Huang, H., Lin, L., Tong, R., Hu, H., Zhang, Q., Iwamoto, Y., Han, X., Chen, Y.W., Wu, J.: Unet 3+: A full-scale connected unet for medical image segmentation. In: ICASSP 2020-2020 IEEE International Conference on Acoustics, Speech and Signal Processing (ICASSP). pp. 1055--1059. IEEE (2020)

\bibitem{jha2020kvasir}
Jha, D., Smedsrud, P.H., Riegler, M.A., Halvorsen, P., de~Lange, T., Johansen, D., Johansen, H.D.: Kvasir-seg: A segmented polyp dataset. In: Proceedings of the 26th International Conference on MultiMedia Modeling. pp. 451--462 (2020)

\bibitem{kawaguchi2017generalization}
Kawaguchi, K., Kaelbling, L.P., Bengio, Y.: Generalization in deep learning. arXiv preprint arXiv:1710.05468  \textbf{1}(8) (2017)

\bibitem{kumar2019multi}
Kumar, N., Verma, R., Anand, D., Zhou, Y., Onder, O.F., Tsougenis, E., Chen, H., Heng, P.A., Li, J., Hu, Z., et~al.: A multi-organ nucleus segmentation challenge. IEEE transactions on medical imaging  \textbf{39}(5),  1380--1391 (2019)

\bibitem{li2023global}
Li, P., Zhou, R., He, J., Zhao, S., Tian, Y.: A global-frequency-domain network for medical image segmentation. Computers in Biology and Medicine p. 107290 (2023)

\bibitem{mendoncca2013ph}
Mendon{\c{c}}a, T., Ferreira, P.M., Marques, J.S., Marcal, A.R., Rozeira, J.: Ph 2-a dermoscopic image database for research and benchmarking. In: 2013 35th annual international conference of the IEEE engineering in medicine and biology society (EMBC). pp. 5437--5440 (2013)

\bibitem{morrison2021exploring}
Morrison, K., Gilby, B., Lipchak, C., Mattioli, A., Kovashka, A.: Exploring corruption robustness: inductive biases in vision transformers and mlp-mixers. arXiv preprint arXiv:2106.13122  (2021)

\bibitem{neyshabur2017exploring}
Neyshabur, B., Bhojanapalli, S., McAllester, D., Srebro, N.: Exploring generalization in deep learning. Advances in neural information processing systems  \textbf{30} (2017)

\bibitem{park2022vision}
Park, N., Kim, S.: How do vision transformers work? arXiv preprint arXiv:2202.06709  (2022)

\bibitem{raghu2021vision}
Raghu, M., Unterthiner, T., Kornblith, S., Zhang, C., Dosovitskiy, A.: Do vision transformers see like convolutional neural networks? Advances in Neural Information Processing Systems  \textbf{34},  12116--12128 (2021)

\bibitem{wang2023theoretical}
Wang, Z., Wu, L.: Theoretical analysis of inductive biases in deep convolutional networks. arXiv preprint arXiv:2305.08404  (2023)

\bibitem{wei2022masked}
Wei, C., Fan, H., Xie, S., Wu, C.Y., Yuille, A., Feichtenhofer, C.: Masked feature prediction for self-supervised visual pre-training. In: Proceedings of the IEEE/CVF Conference on Computer Vision and Pattern Recognition. pp. 14668--14678 (2022)

\bibitem{yang2022source}
Yang, C., Guo, X., Chen, Z., Yuan, Y.: Source free domain adaptation for medical image segmentation with fourier style mining. Medical Image Analysis  \textbf{79},  102457 (2022)

\bibitem{yang2020fda}
Yang, Y., Soatto, S.: Fda: Fourier domain adaptation for semantic segmentation. In: Proceedings of the IEEE/CVF conference on computer vision and pattern recognition. pp. 4085--4095 (2020)

\end{thebibliography}
